\documentclass{elsart}%
\usepackage{graphicx,amssymb,amsmath,bm}
\newcommand{\ep}{\epsilon}
\begin{document}
%\begin{titlepage}
% \vskip 0.3cm
%\begin{center}
%{\Large\bf Differential equations and high-energy expansion of two--loop
%diagrams in $D$ dimensions}
%\end{center}
%\vskip 0.5cm \centerline{A.V.~Bogdan$^{\dag}$ and R.N.~Lee$^{\ddag}$} \vskip .3cm
%\begin{center}
%{\sl Institute of Nuclear Physics, 630090 Novosibirsk, Russia}
%\end{center}
%\vskip 1cm
%\begin{abstract}
%New method of calculation of master integrals using differential equations and
%asymptotical expansion is presented. This method leads to the results exact in
%space-time dimension $D$ having the form of the convergent power series. As an
%application of this method, we calculate the two--loop master integral for
%"crossed--triangle" topology which was previously known only up to $O(\ep)$
%order. The case when a topology contains several master integrals is also
%considered. We present an algorithm of the term-by-term calculation of the
%asymptotical expansion in this case and analyze in detail the "crossed--box"
%topology with three master integrals.
%\end{abstract}
%%\vskip .5cm
%\vfill \hrule \vskip.3cm \noindent $^{\ast}${\it Work supported by the Russian Fund of
%Basic Researches, project 03-02-16529-a.} \vspace{1cm}\\ $
%\begin{array}{ll} ^{\dag}\mbox{{\it e-mail address:}} &
%\mbox{A.V.Bogdan@inp.nsk.su}\\
%^{\ddag}\mbox{{\it e-mail address:}} &
%\mbox{R.N.Lee@inp.nsk.su}\\
%\end{array}
%$
%\end{titlepage}
\begin{frontmatter}
\title{Differential equations and high-energy expansion of two--loop
diagrams in $D$ dimensions\thanksref{RFBR}}
\thanks[RFBR]{Work supported by the Russian Fund of Basic
Researches, project 03-02-16529-a.}
\author{A.V.~Bogdan}
\ead{A.V.Bogdan@inp.nsk.su}
 and \author{R.N.~Lee}\ead{R.N.Lee@inp.nsk.su}
\address{Budker Institute of Nuclear Physics, 630090 Novosibirsk, Russia}

\begin{abstract}
New method of calculation of master integrals using differential equations and asymptotical
expansion is presented. This method leads to the results exact in space-time dimension $D$ having
the form of the convergent power series. As an application of this method, we calculate the
two--loop master integral for "crossed--triangle" topology which was previously known only up to
$O(\ep)$ order. The case when a topology contains several master integrals is also considered. We
present an algorithm of the term-by-term calculation of the asymptotical expansion in this case and
analyze in detail the "crossed--box" topology with three master integrals.
\end{abstract}
\end{frontmatter}

\section{Introduction}

High precision of  modern high-energy experiments stimulates interest to two--loop
calculations. These calculations, in comparison with one-loop calculations, are technically
much more complicated. The tensor reduction and integration-by-parts (IBP) method
\cite{Chetyrkin_1,Laporta_1} effectively reduces the integral, corresponding to any multi-loop
diagram, to the linear combination of the master integrals (MIs). The coefficients in this
linear combination often have poles of some finite order at $\epsilon=0$ (space-time dimension
is $D=4-2\epsilon$). Therefore, it is necessary to know the expansion of the master integrals
in $\epsilon$, at least, a few first terms of this expansion. Successful method of calculating
this expansion term-by-term is based on the application of the differential equations method
\cite{Kotikov_1,Remiddi2000,Remiddi2001,Birthwright2004}. The results for two-loop diagrams
are usually expressed in terms of (generalized) harmonic polylogarithms
\cite{Remiddi2001,Birthwright2004} (see, however, Ref. \cite{Remiddi2000}). Due to rapid
increase of calculational complexity when increasing the order of expansion, the calculations
using this method are typically done up to $O(\epsilon)$ terms. However, it may be not
sufficient for the calculation of some diagrams. Therefore, it is highly desirable to
calculate master integrals exactly in $D$. One of the most powerful methods of such
calculation is the application of Mellin--Barnes transformation \cite{Boos1991,Friot2005}.
However, for the case of several external invariants and/or many internal lines this method
becomes hard to apply.

In this paper, we present a method of obtaining the asymptotic expansion (AE) in inverse powers of
some large scale $s$ of multi-loop master integrals exact in space--time dimension $D$. Moreover,
this expansion has a finite radius of convergence which can be determined from ODE theory, thus
being a power series representation accessible for numerical calculations in definite kinematic
region. As an application, we calculate series representations of two--loop diagram presented in
Fig.\ref{Marquee} in two different kinematics. Our result can be easily expanded in $\epsilon$ and
the $O(\epsilon^0)$ term reproduces the known result. We also present the algorithm allowing to
calculate, term--by--term, the asymptotical expansion of the diagrams in Fig.\ref{Diamond}.

Our method can be described as follows. For the calculation of the master integrals of given
topology we differentiate them with respect to $s$, and express the result of differentiation
(with help of integration by parts \cite{Chetyrkin_1,Laporta_1}) in terms of MIs of the same
topology and its subtopologies. Thus, we obtain, in general case, the linear system of
first--order ODEs. We assume that the MIs of subtopologies are already known by means of this
method or some other. When the given topology has only one MI, $j$, the system is reduced to
one equation. The authors of papers \cite{Remiddi2001,Birthwright2004} also used the
differential equations method and searched for the solution in the form of
$\epsilon$--expansion imposing the boundary condition at some finite point. Unfortunately, the
results obtained by the method proposed in Refs. \cite{Remiddi2001,Birthwright2004} quickly
become very cumbersome with increase of the order in $\epsilon$. The original point of our
approach is the choice of boundary conditions. We impose boundary condition by fixing a
definite term of asymptotical expansion of $j$, namely, the term proportional to the main
asymptotics of the solution of the homogeneous part of the differential equation, $j_h$. In
the cases considered in this paper, choice $\epsilon>0$ ($D=4-2\epsilon$) allows to fix the
boundary condition by calculation of the main asymptotics of $j$. We calculate this
asymptotics by means of standard techniques (Feynman parametrization, Mellin--Barnes
representation, etc.). Given the asymptotical expansion of inhomogeneous term, of $j_h$, and
boundary condition, we find the asymptotical expansion of $j$. From the standard theory of ODE
it follows that obtained power series has a finite radius of convergence.

Our approach can be also used in the case of system of several equations. In
this case, the additional obstacles connected with finding the homogeneous
solution arise. However, we can find the AE of the  homogeneous solution up to
any finite order. The same is true for solution of initial inhomogeneous
system.

Thus, using our approach, one can reduce the calculation of the AE (having the form of covergent
power series) of master integral in $D$ dimensions to the calculation of its main asymptotics. The
radius of convergence can be determined by the methods of ODE theory.

The remainder of the paper is organized as follows. In Section 2, we consider
the case of one MI in the topology. We calculate the AE of the integrals
corresponding to the vertex diagrams shown in Fig.\ref{Marquee}. In Section 3,
we describe the algorithm which allows to calculate the AE of MIs in the case
of system of differential equations. As an example, we consider the
crossed--box topology with three MIs depicted in Fig.\ref{Diamond}. In the
Conclusion, we discuss the main results of the present paper.

\section{Topology with one master integral}

Consider inhomogeneous differential equation for the MI $j$ with respect to
external scale $s$
\begin{equation}\label{1}
    \frac{d}{ds}j(s)=f(s)j(s)+h(s)\,.
\end{equation}
The solution of this equation is
\begin{equation}\label{4}
  j(s)=j_h(s)\int^s_{s_0}j^{-1}_h(s')h(s')ds'+j_h(s)j_h^{-1}(s_0)j(s_0)\,,
\end{equation}
where $j_h(s)=\exp[\int f(s)ds]$ is the solution of  the corresponding
homogeneous equation.

\begin{figure}
\centering
\[
F_1(s,p^2,q^2)=\hspace{-1cm}\raisebox{-48\unitlength}[48\unitlength]{
\includegraphics{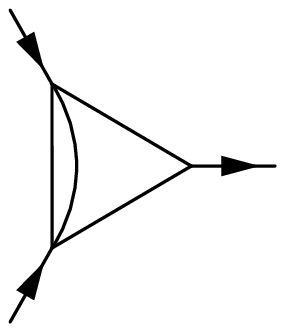}
\begin{picture}(0,0)(105,0)
\put(24.29472,81.4194){\makebox(0,0)[rt]{$q$}}%
\put(34.73712,12.66714){\makebox(0,0)[lt]{$p$}}%
\put(87.91696,56){\makebox(0,0)[b]{$p+q$}}%
\end{picture}
},\qquad
%%%%
F_2(s,p^2,q^2)=\hspace{-1cm}\raisebox{-48\unitlength}[48\unitlength]{
\includegraphics{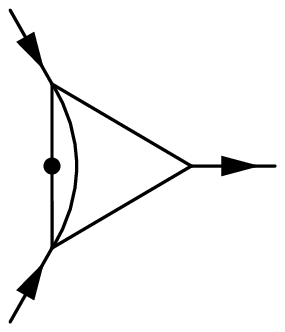}
\begin{picture}(0,0)(105,0)
\put(24.29472,81.4194){\makebox(0,0)[rt]{$q$}}%
\put(34.73712,12.66714){\makebox(0,0)[lt]{$p$}}%
\put(87.91696,56){\makebox(0,0)[b]{$p+q$}}%
\end{picture}
},\quad\quad G(p^2)=\hspace{-0.2cm}\raisebox{-48\unitlength}[48\unitlength]{
\includegraphics{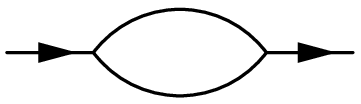}
\begin{picture}(0,0)(105,0)
\put(12.50002,44){\makebox(0,0)[t]{$p$}}%
\put(87.50002,44){\makebox(0,0)[t]{$p$}}%
\end{picture}
}
\]
\caption{Master integrals entering Eq.\eqref{de1}.} \label{inhom1}
\end{figure}

In our approach we are interested in the asymptotical expansion for large $s$ and assume that it is
already known for $h(s)j^{-1}_h(s)$. Then it is convenient to let $s_0$ tend to infinity:
\begin{equation}\label{5}
  j(s)=j_h(s)\left.\left(\int^s_{s_0}j^{-1}_h(s')h(s')ds'+
  j_h^{-1}(s_0)j(s_0)\right)\right|_{s_0\rightarrow \infty}\,.
\end{equation}
For MIs considered in the present paper, the choice $\epsilon>0$ allows to calculate the limit
$s_0\to \infty$ separately in the first and the second terms in Eq. \eqref{5}. In this case we can
integrate term-by-term the AE of the integrand in Eq. \eqref{5}. In this stage the problem of
finding of the whole AE of $j$ is reduced to the calculation of its main asymptotics. In the more
general case, when the limits can not be taken separately, the problem is reduced to the
calculation of few first terms of the asymptotics of $j$.
\begin{figure}
\centering
\[
j_1=\hspace{-1.2cm}\raisebox{-60\unitlength}[60\unitlength]{
\includegraphics{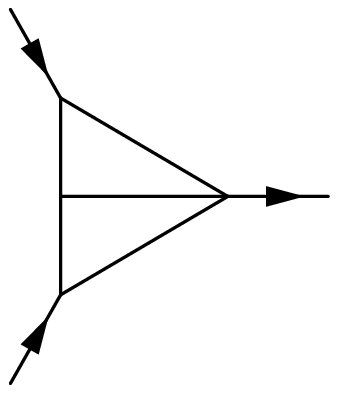}
\begin{picture}(0,0)(125,0)
\put(40.63985,104.20781){\makebox(0,0)[lb]{$q$}}%
\put(40.64035,15.7919){\makebox(0,0)[lt]{$p$}}%
\put(105.50035,66){\makebox(0,0)[b]{$p+q$}}%
\end{picture}}\qquad
%%%%
j_2=\hspace{-1.2cm}\raisebox{-60\unitlength}[60\unitlength]{\includegraphics{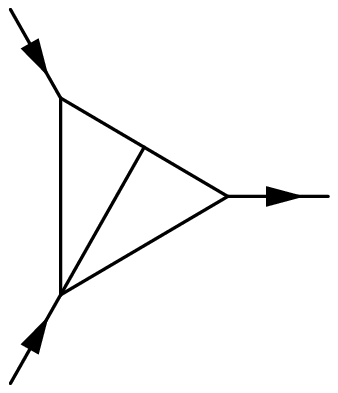}
\begin{picture}(0,0)(125,0)
\put(40.63985,104.20781){\makebox(0,0)[lb]{$q$}}%
\put(40.64035,15.7919){\makebox(0,0)[lt]{$p$}}%
\put(105.50035,66){\makebox(0,0)[b]{$p+q$}}%
\end{picture}
 }
\]
\caption{"Crossed--triangle" topology.} \label{Marquee}
\end{figure}
As an application of our method, we calculate the AEs of the master integral
$j_1$ shown in Fig. \ref{Marquee} in two different kinematics. The analytical
expression corresponding to the diagram in Fig. \ref{Marquee}a has the form
\begin{equation}
j_1(s,p^2,q^2)=\frac{1}{(2\pi)^{2D}}\int\frac{d^D r d^D l}{r^2
    l^2(r-p)^2(l-q)^2(r+l)^2}\,,
\end{equation}
where $s=(p+q)^2$ and the usual prescription $r^{2}\to r^{2}+i0$ is implied.
For convenience, we assume that $p^2,q^2,s<0$, and the final result should be
understood as the analytical continuation from this region. For the kinematics
presented in Fig. \ref{Marquee}a the differential equation has the form
\begin{equation}\label{de1}
   \frac{\partial j_1(s,p^2,q^2)}{\partial s}=\frac{(3-D)y}{2\lambda^2}
   j_1(s,p^2,q^2)+\frac{1}{2\lambda^2}h(s,p^2,q^2)\,,
\end{equation}
where
\begin{equation}\label{H}
\begin{split}
   h(s,p^2,q^2)&=-(y+p^2)F_2(q^2,p^2,s)-(y+q^2)F_2(p^2,q^2,s)\\
   \mbox{}&+\frac{(3-D)(3D-10)}{D-4}\left(F_1(q^2,p^2,s)
    +F_1(p^2,q^2,s)\right)\\
    \mbox{}&+\frac{2(3-D)^2}{D-4}G(p^2)G(q^2)\,,
\end{split}
\end{equation}
and $y=(p\cdot q)=(s-p^2-q^2)/2$, $\lambda^2=y^2-p^2q^2$, and
$F_1(p_1^2,p_2^2,p_3^2)$, $F_2(p_1^2,p_2^2,p_3^2)$, and $G(p_1^2)$ are given by
three-- and two--point integrals presented in Fig. \ref{inhom1} (dot
corresponds to the square of denominator). The asymptotical expansion for $F_1$
and $F_2$ can be obtained from the well--known formula for off--shell massless
triangle with arbitrary powers of denominators \cite{Boos1991}. This expansion
and formula for $G(p_1^2)$ are presented in the Appendix.

The homogeneous solution of Eq. \eqref{de1} is
\begin{equation}\label{hs1}
    j_{1h}(s,p^2,q^2)=(-2y)^{-1+2\epsilon}\left(1-\frac{p^2q^2}{y^2}\right)^{-1/2+\epsilon}
    \stackrel {s\to\infty}\longrightarrow (-s)^{-1+2\epsilon}\,.
\end{equation}

The choice $\epsilon>0$ allows us to take the limit separately in the first and the second terms in
Eq. \eqref{5}. Indeed, the leading asymptotics of $h(s,p^2,q^2)$ is $s^{\epsilon}$. Thus, the
integrand in Eq. \eqref{5} behaves as $s^{-1-\epsilon}$ and the integral converges when its lower
limit is replaced by $\infty$. Therefore, we can confine ourselves to calculation of the main
asymptotics of $j_{1}$ that determines the limit
\begin{equation}
\left.\left(j_{1h}^{-1}(s_0,p^2,q^2)j_1(s_0,p^2,q^2)\right)\right|_{s_0\rightarrow
\infty}.
\end{equation}

We calculate this asymptotics with the help of Feynman parametrization
\begin{equation}\label{fp1}
    j_1(s,p^2,q^2)=
\frac{\Gamma[1+2\epsilon]}{(4\pi)^D}\int_0^\infty\frac{\prod_{i=1}^5dx_i\,
\delta(1-x_{12345})(x_{13}x_{24}+x_5x_{1234})^{-1+3\epsilon}}{(-s
x_3x_4x_5-p^2x_3(x_5x_{12}+x_1x_{24})-q^2x_4(x_5x_{12}+x_2x_{13}))^{1+2\epsilon}}\,.
\end{equation}
where $x_{i_1,i_2,\ldots}=x_{i_1}+x_{i_2}+\cdots$. The region, providing the
main asymptotics, is
\begin{equation}\label{re1}
0\leqslant x_3\sim x_4\ll x_1 \sim x_2\sim x_5 \leqslant 1\,.
\end{equation}
The result of integration is
\begin{equation}\label{main_as}
    j_1(s,p^2,q^2)\stackrel {s\to\infty}\longrightarrow\frac{\pi^2}{(4\pi)^D}
    \frac{\Gamma[2\epsilon]}{\epsilon\sin(\pi\ep)\sin(2\pi\ep)}
    \frac{(-s)^{-1+2\epsilon}}{(-q^2)^{2\epsilon}(-p^2)^{2\epsilon}}\,.
\end{equation}
We use the following expansion
\begin{equation}\label{lambda_expansion}
    \lambda^{-2}j^{-1}_{1h}(s,p^2,q^2)=
    (-s)^{-1-2\epsilon}\frac{2\Gamma[\epsilon]}{\Gamma[2\epsilon]}
    \sum_{k,l=0}^{\infty}
    \Gamma\left[\substack{\displaystyle k+l+1+2\epsilon, k+l+1+\epsilon\\
                    \displaystyle k+1+\epsilon,l+1+\epsilon}\right]
    \frac{(p^2/s)^k(q^2/s)^l}{k!\;l!}\,,
\end{equation}
where $\Gamma\left[\substack{x_1,x_2,\ldots\\
 y_1,y_2,\ldots}\right]=
 \frac{\Gamma[x_1,x_2,\ldots]}{\Gamma[y_1,y_2,\ldots]}=
 \frac{\Gamma[x_1]\Gamma[x_2]\cdots}{\Gamma[y_1]\Gamma[y_2]\cdots}$ .
Substituting \eqref{lambda_expansion} in Eq. \eqref{5} and integrating over $s$, we obtain
\begin{equation}\label{j1_ans}
\begin{split}
j_1&=(-2y)^{-1+2\ep}\left(1-\frac{p^2q^2}{y^2}\right)^{-1/2+\ep}{(4\pi)^{-D}}\\
{}\times&\left\{\Gamma[2\ep]\frac{\pi^2}{\ep\sin(\pi\ep)\sin(2\pi\ep)}(-q^2)^{-2\ep}(-p^2)^{-2\ep}\right.\\
{}+&4\pi^2\cot^2(\pi \ep)(-q^2)^{-\ep}(-p^2)^{-\ep}(-s)^{-2\ep}
\Gamma[2\ep]\sum_{k,l=0}^{\infty}\frac{(p^2/s)^k(q^2/s)^l}{k!\;l!}\Gamma\left[\substack{\displaystyle
k+l+2\ep, k+l+1+\ep\\\displaystyle 1+\ep,k+1+\ep,l+1+\ep}\right]\\
{}-&\frac{\pi^3}{2\Gamma[-3\ep]\sin^3(\pi \ep)}
\sum_{k,l,m,n=0}^\infty\frac{(p^2/s)^{k+m}(q^2/s)^{l+n}}{k!\:l!\:m!\:n!}
\Gamma\left[\substack{\displaystyle
k+l+1+2\ep, k+l+1+\ep\\\displaystyle 1+\ep, k+1+\ep,l+1+\ep}\right]\\
{}\times&\left[\frac{(-s)^{-4\ep}}{k+l+m+n+4\ep}\left(m\:\Gamma\left[\substack{\displaystyle m+n,m+n+2\ep\\
\displaystyle
1+\ep+m,1+2\ep+n}\right]+\frac{\delta_{n+m=0}}{3\ep\Gamma[1+\ep]}\right)\right.\\
{}&-\frac{(-s)^{-2\ep}(-q^2)^{-2\ep}}{k+l+m+n+2\ep}\left(m\:\Gamma\left[\substack{\displaystyle m+n,m+n-2\ep\\
\displaystyle
1+\ep+m,1-2\ep+n}\right]-\frac{\delta_{n+m=0}}{3\ep\Gamma[1+\ep]}\right)\\
{}&-\frac{(-s)^{-3\ep}(-p^2)^{-\ep}}{k+l+m+n+3\ep}\Gamma\left[\substack{\displaystyle m+n+\ep,m+n-\ep\\
\displaystyle
-\ep+m,1+2\ep+n}\right]\\
{}&+\frac{(-s)^{-\ep}(-p^2)^{-\ep}(-q^2)^{-2\ep}}{k+l+m+n+\ep}\Gamma\left[\substack{\displaystyle m+n-\ep,m+n-3\ep\\
\displaystyle -\ep+m,1-2\ep+n}\right]+\left(\substack{\displaystyle
p\leftrightarrow q\\
\displaystyle m\leftrightarrow n}\right)\biggr]\biggr\}\,.
\end{split}
\end{equation}
Then we make a shift $k\to k-m$, $l\to l-n$ and change the summation order. The
finite sums over $m,n$ can be expressed in closed forms. The final result reads
\begin{equation}\label{j1_answer}
\begin{split}
j_1&=(-2y)^{-1+2\ep}\left(1-\frac{p^2q^2}{y^2}\right)^{-1/2+\ep}{(4\pi)^{-D}}\\
&{}\times\left\{\Gamma[2\ep]\frac{\pi^2}{\ep\sin(\pi\ep)\sin(2\pi\ep)}(-q^2)^{-2\ep}(-p^2)^{-2\ep}\right.\\
&{}+4\Gamma[2\ep]\pi^2\cot^2(\pi
\ep)(-q^2)^{-\ep}(-p^2)^{-\ep}(-s)^{-2\ep}
\sum_{k,l=0}^{\infty}\frac{(p^2/s)^k(q^2/s)^l}{k!\;l!}\Gamma\left[\substack{\displaystyle
l+k+2\ep, l+k+1+\ep\\\displaystyle 1+\ep,l+1+\ep,k+1+\ep}\right]
\\
&{}+\ep^2\Gamma[-\ep]^3\Gamma[\ep]%
\sum_{k,l=0}^{\infty}% sum sum sum sum sum sum
\frac{(p^2/s)^k(q^2/s)^l}{k!\;l!} %
\\
&{}\times\biggl\{-(-s)^{-4\ep}
\Gamma\left[\substack{\displaystyle k+l+4\ep,k+l+1+2\ep\\
\displaystyle 1-3\ep,l+1+4\ep,1+k+\ep}\right]\frac1{2\ep}
{}_3F_2\left[\substack{\displaystyle -k,3\ep ,2\ep
\\\displaystyle 1+l+4\ep,1+2\ep}\bigr| 1\right]\\
&{}-(-s)^{-2\ep}(-q^2)^{-2\ep}
\Gamma\left[\substack{\displaystyle l+k+1-2\ep,k+l+2\ep\\
\displaystyle 1-3\ep,l+1-2\ep,1+k+\ep}\right]\frac1{2\ep}
{}_3F_2\left[\substack{\displaystyle -k,3\ep ,2\ep
\\\displaystyle -l-k+2\ep ,1+2\ep}\bigr| 1\right]\\
&{}+(-s)^{-3\ep}(-p^2)^{-\ep}
\Gamma\left[\substack{\displaystyle 2\ep,\ep,k+l+3\ep,k+l+1+\ep\\
\displaystyle 1-3\ep,3\ep,1+l+2\ep,1+k+\ep}\right]\\
&+(-s)^{-\ep}(-p^2)^{-\ep}(-q^2)^{-2\ep}
\Gamma\left[\substack{\displaystyle 2\ep,k+l+1-\ep,k+l+\ep\\
\displaystyle
1-\ep,1+l-2\ep,1+k+\ep}\right]+\left(\substack{\displaystyle
p\leftrightarrow q\\
\displaystyle k\leftrightarrow l}\right)\biggr\}\biggr\} \,.
\end{split}
\end{equation}
As known, $j_1$ is finite at $D=4$. It is easy to check that $\epsilon$-poles
in individual terms cancel and the finite part reproduce the AE of the
well-known result of \cite{Usyukina1994}.

Let us consider the homogeneous system of the differential equations in $s$ for
all MIs appearing in the subtopologies of $j_1$. The point $s=\infty$ is a
regular singular point of this system. Then, from the theory of ODE, it follows
that the above expansion has a finite radius of convergence, determined by the
closest singularity of the coefficients of this system. In our case, it is
located in the point $\lambda^2=0$. Thus, the convergence region of Eq.
\eqref{j1_answer} is determined by the condition
\begin{equation}\label{convergence_radius}
\lvert s\rvert > \lvert \sqrt{-p^2}\pm \sqrt{-q^2}\rvert^{2}\,.
\end{equation}

%%%%%%%%%%%%%%%%%%%%%%%%%%%%%%%%%%%%%%%%%%%%%%%%%%%%%%%%%%%%%%%%%%%%%%%
%%%%%%%%%%%%%%%%%%%%%%%%%%%%%%%%%%%%%%%%%%%%%%%%%%%%%%%%%%%%%%%%%%%%%%%
%%%%%%%%%%%%%%%%%%%%%%%%%%%%%%%%%%%%%%%%%%%%%%%%%%%%%%%%%%%%%%%%%%%%%%%
%%%%%%%%%%%%%%%%%%%%%%%%%%%%%%% j2 %%%%%%%%%%%%%%%%%%%%%%%%%%%%%%%%%%%%
%%%%%%%%%%%%%%%%%%%%%%%%%%%%%%%%%%%%%%%%%%%%%%%%%%%%%%%%%%%%%%%%%%%%%%%
%%%%%%%%%%%%%%%%%%%%%%%%%%%%%%%%%%%%%%%%%%%%%%%%%%%%%%%%%%%%%%%%%%%%%%%
%%%%%%%%%%%%%%%%%%%%%%%%%%%%%%%%%%%%%%%%%%%%%%%%%%%%%%%%%%%%%%%%%%%%%%%
%%%%%%%%%%%%%%%%%%%%%%%%%%%%%%%%%%%%%%%%%%%%%%%%%%%%%%%%%%%%%%%%%%%%%%%

Now we pass to the calculation of the asymptotical expansion at large $s$ of master integral
$j_2(s,p^2,q^2)$ corresponding to the diagram shown in Fig. \ref{Marquee}b,
\begin{equation}
j_2(s,p^2,q^2)=j_1(p^2,s,q^2)\,.
\end{equation}
The differential equation has the form
\begin{equation}\label{de2}
\frac{\partial j_2(s,p^2,q^2)}{\partial
s}=\left(\frac{D-4}{s}-\frac{(D-3) y}{2\lambda ^2}\right)
j_2(s,p^2,q^2)+\frac{1}{2\lambda ^2}h(s,p^2,q^2)\,,
\end{equation}

where
\begin{equation}\label{inhom2}
\begin{split}
   h(s,p^2,q^2)&=p^2 F_2(q^2,p^2,s)-\frac{p^2}{s}(q^2+y) F_2(s,p^2,q^2)\\
   &{}+\frac{(3D-10)(D-3)}{s(D-4)}(p^2+y)(F_1(q^2,p^2,s)+F_1(s,p^2,q^2))\\
   &{}-2\frac{(D-3)^2}{s(D-4)}(p^2+y)G(q^2)G(s)\,.
\end{split}
\end{equation}

The homogeneous solution of Eq. \eqref{de2} is
\begin{equation}\label{hs2}
    j_{2h}(s,p^2,q^2)=(-s)^{-2\epsilon}(-2y)^{-1+2\epsilon}
    \left(1-\frac{p^2q^2}{y^2}\right)^{-1/2+\epsilon}
    \stackrel {s\to\infty}\longrightarrow (-s)^{-1}\,.
\end{equation}

Similar to the previous case, the choice $\epsilon>0$ allows us to take the
limit separately in the first and the second terms in Eq.
\eqref{5}. The calculation of the main asymptotics of
$j_2(s,p^2,q^2)$ is slightly more difficult than that for $j_1(s,p^2,q^2)$.
Using the parametric representation we have
\begin{equation}\label{fp2}
j_2(s,p^2,q^2)=
\frac{\Gamma[1+2\epsilon]}{(4\pi)^D}\int_0^\infty\frac{\prod_{i=1}^5dx_i\,
\delta(1-\sum x_i)(x_{13}x_{24}+x_5x_{1234})^{-1+3\epsilon}}{(-p^2
x_3x_4x_5-sx_3(x_5x_{12}+x_1x_{24})-q^2x_4(x_5x_{12}+x_2x_{13}))^{1+2\epsilon}}\,.
\end{equation}
As it is well-known, the sum in the argument of the $\delta$-function in Eq.
\eqref{fp2} can run over arbitrary subset of $x_i$. We choose it as $\sum
x_i=x_{1235}$. The region, providing the main asymptotics, is
\begin{equation}\label{re2}
0\leqslant x_3\ll x_1 \sim x_2\sim x_4\sim x_5 \lesssim 1\,.
\end{equation}
After the integration over $x_3$ we obtain
\begin{equation}
j_2(s,p^2,q^2)\stackrel {s\to\infty}\longrightarrow
\frac{\Gamma[2\epsilon]}{(4\pi)^D}\frac{(-q^2)^{-2\epsilon}}{-s}\int_0^\infty \frac{dx_1 dx_2 dx_4
dx_5 \delta(1-x_{125})\beta^{-2\epsilon}x_4^{-2\epsilon}}
{(\beta+x_4x_1)(\beta+x_4x_{15})^{1-3\epsilon}}\,,
\end{equation}
where $\beta=x_1x_2+x_2x_5+x_5x_1$. Now we use the Mellin--Barnes
representation
\begin{equation}
\frac1{(\beta+x_4x_{15})^{1-3\epsilon}}=\frac1{\Gamma[1-3\epsilon]}
\int\limits_{-i\infty+0}^{i\infty+0}\frac{ds}{2\pi i}
\frac{\Gamma[1-3\epsilon-s]}{\beta^{1-3\epsilon-s}}\frac{\Gamma[s]}{(x_4x_{15})^{s}}\,,
\end{equation}
and take the integrals over $x_i$. The final integration over $s$ can be done
by applying the Barnes second lemma (see, e.g., \cite{Slater1966}) and results
in
\begin{equation}
j_2(s,p^2,q^2)\stackrel {s\to\infty}\longrightarrow
\frac{1}{(4\pi)^D}\frac{(-q^2)^{-2\epsilon}}{-s}
\frac{\Gamma[1-\epsilon]^3\Gamma[2\epsilon]}{\epsilon(1-2\epsilon)^2\Gamma[1-3\epsilon]}
\,{}_3F_2\left[\substack{\displaystyle 1,1,1-\epsilon
\\\displaystyle 2-2\epsilon,2-2\epsilon}\bigr| 1\right]
\end{equation}
The final result for the AE of $j_2(s,p^2,q^2)$, for the diagram depicted in
Fig.\ref{Marquee}b, reads
\begin{equation}\label{j2_answer}
\begin{split}
j_2=&\frac{(-2y)^{-1+2\ep}(-s)^{-2\ep}}{(4\pi)^{D}}\left(1-\frac{p^2q^2}{y^2}\right)^{-1/2+\ep}\left\{
(-q^2)^{-2\ep}\frac{\Gamma[1-\ep]^3\Gamma[2\ep]}{\ep(1-2\ep)^2\Gamma[1-3\ep]}\;\:{_3\mathrm{F}_2}
\left[
\substack{\displaystyle %
1,\;1,\;1-\ep\\ \displaystyle
2-2\ep,2-2\ep
}\bigr|1\right]\right.\\
{}+&\frac{\Gamma[\ep]}{\Gamma[2\ep]}
\sum_{k,l=0}^\infty\left(\frac{p^2}{s}\right)^k
\left(\frac{q^2}{s}\right)^l\left[
(-q^2)^{-\ep}(-s)^{-\ep}\frac{\Gamma[1-\ep]^4\Gamma[\ep]^2}{2\ep\Gamma[1-2\ep]^2}\;B_{k,l}\right.\\
{}+&\frac{\Gamma[1-\ep]^2}{2\ep\Gamma[1-3\ep,1-2\ep]}\left(\frac{}{}
(-q^2)^{-\ep}(-s)^{-\ep}C^{(1)}_{k,l}+(-p^2)^{-2\ep}(-q^2)^{-\ep}(-s)^{\ep}C^{(2)}_{k,l}
+(-p^2)^{-2\ep}(-q^2)^{-2\ep}(-s)^{2\ep}C^{(3)}_{k,l}\right.\\
{}&+(-s)^{-2\ep}(C^{(4)}_{k,l}+C^{(5)}_{k,l})+(-p^{2})^{-2\ep}(C^{(6)}_{k,l}+C^{(7)}_{k,l})
+(-q^{2})^{-2\ep}C^{(8)}_{k,l} \left.\left.\left.\frac{}{}\right)\right]\right\}\,,
\end{split}
\end{equation}
\begin{equation}
    A_{k,l,m,n}=\frac{(-1)^{n+m}}{n!m!(k-m)!(l-n)!}
\;\Gamma\left[
\substack{\displaystyle %
k+l-m-n+1+2\ep,k+l-m-n+1+\ep\\ \displaystyle k-m+1+\ep,\;l-n+1+\ep }\right]\,,
\end{equation}
\begin{equation}
    B_{k,l}=\frac{1}{k+l+\ep}(A_{k,l,1,0}\delta_{k>0}-A_{k,l,0,1}\delta_{l>0}-A_{k,l,0,0})\,,
\end{equation}
\begin{equation}
\begin{split}
C^{(1)}_{k,l}=&\sum_{m=0}^k\sum_{n=0}^l A_{k,l,m,n}
\frac{1-\ep}{k+l+\ep}
\Gamma[m+n-\ep,m+n-1+\ep,1-2\ep-m,1+\ep-n]=\mbox{}\\
{}&=\frac{\pi^2}{\sin[2\pi\ep]\sin[\pi\ep]}\frac{\Gamma[2\ep]}{\Gamma[3\ep]}\frac{2}{(k+l+\ep)k!l!}\;
\Gamma\left[
\substack{\displaystyle %
k+l+3\epsilon,k+l+1+\epsilon\\ \displaystyle k+2\epsilon,l+1+\epsilon }\right]\,,
\end{split}
\end{equation}
\begin{equation}
\begin{split}
C^{(2)}_{k,l}=&\sum_{m=0}^k\sum_{n=0}^l A_{k,l,m,n}
\frac{\ep-1}{k+l-\ep}
\,m\,\Gamma[m+n-3\ep,m+n-1-\ep,2\ep-m,1+\ep-n]=\mbox{}\\
{}&=\frac{\pi\Gamma[2\ep,1-3\ep]}{\sin[2\pi\ep]}\;\frac{2k}{(k+l-\ep)k!l!}\; \Gamma\left[
\substack{\displaystyle %
k+l+\epsilon,k+l+1-\epsilon\\ \displaystyle k+1-2\epsilon,l+1+\epsilon }\right]\,,
\end{split}
\end{equation}
\begin{equation}
\begin{split}
C^{(3)}_{k,l}=&\sum_{m=0}^k\sum_{n=0}^l A_{k,l,m,n}
\frac{3m\ep-n(1-\ep)-6\ep^2}{k+l-2\ep}\,m\,
\Gamma[m+n-2\ep,m+n-1-3\ep,2\ep-m,2\ep-n]=\mbox{}\\
{}&=\frac{\pi \sin[\pi \ep]}{\sin^2[2\pi\ep]}\Gamma[2\ep,1-3\ep] \frac{-2}{(k+l-2\ep)k!l!}\;k\,
\Gamma\left[
\substack{\displaystyle %
k+l-\epsilon,k+l+1-2\epsilon\\ \displaystyle k+1-2\epsilon,l+1-2\epsilon }\right]\,,
\end{split}
\end{equation}
\begin{equation}
\begin{split}
C^{(4)}_{k,l}=&\sum_{m=0}^k\sum_{n=0}^l
\frac{A_{k,l,m,n}}{k+l+2\ep}\left((\ep-1)\,n\,
\Gamma[m+n,m+n-1+2\ep,1-2\ep-m,-\ep-n]\delta_{n>0}+\frac{}{}\right.\\
{}&+\left.\frac{}{}\Gamma[1-2\ep,-\ep,2\ep]\delta_{m+n=0}\right)=\mbox{}\\
{}&=\frac{\pi}{\sin[2\pi\ep]}\Gamma\left[
\substack{\displaystyle %
k+l+1+\epsilon,k+l+1+2\epsilon\\ \displaystyle k+1+\epsilon,l+1+\epsilon }\right]
\frac{\Gamma[-\ep]}{(k+l+2\ep)k!l!}\;{_3\mathrm{F}_2}\left[
\substack{\displaystyle %
1,-l,1-\epsilon\\ \displaystyle 1+k+\epsilon,1+2\epsilon }\bigr|1\right]\,,
\end{split}
\end{equation}
\begin{equation}
\begin{split}
C^{(5)}_{k,l}=&\sum_{m=0}^k\sum_{n=0}^l A_{k,l,m,n}
\frac{n(1-\ep)-3m\ep+2\ep(1-\ep)}{k+l+2\ep}\Gamma[m+n+2\ep,m+n-1+\ep,1-2\ep-m,-2\ep-n]
=\mbox{}\\
{}&=\frac{\pi^2}{\sin^2[2\pi\ep]}\frac{\Gamma[2\ep]}{\Gamma[3\ep]}
\frac{2}{(k+l+2\ep)k!l!}\;\Gamma\left[
\substack{\displaystyle %
k+l+3\epsilon,k+l+1+2\epsilon\\ \displaystyle k+2\epsilon,l+1+2\epsilon }\right]\,,
\end{split}
\end{equation}
\begin{equation}
\begin{split}
C^{(6)}_{k,l}=&\sum_{m=0}^k\sum_{n=0}^l
\frac{A_{k,l,m,n}}{k+l}\left[\frac{}{}(1-\ep)\,m\,n\,
\Gamma[m+n-2\ep,m+n-1,2\ep-m,-\ep-n]\delta_{m+n>1}\right.\\
{}&-\left.m\,\Gamma[1-2\ep,-\ep,2\ep]\delta_{m+n=1}\frac{}{}\right]
=\mbox{}\\
{}&=\frac{\pi}{\sin[2\pi\ep]}\frac{\Gamma[1-\ep]}{\ep} \frac{-k}{(k+l)k!l!}\;\Gamma\left[
\substack{\displaystyle %
k+l+2\epsilon,k+l+1\\ \displaystyle k+1,l+1+\epsilon }\right]{_3\mathrm{F}_2} \left[
\substack{\displaystyle %
-l,2\ep,1-\ep\\ \displaystyle k+1,1+2\ep }\bigr|1\right]\delta_{l+k>0}\,,
\end{split}
\end{equation}
\begin{equation}
\begin{split}
C^{(7)}_{k,l}=&\sum_{m=0}^k\sum_{n=0}^l
\frac{A_{k,l,m,n}}{k+l}(3m\ep-n(1-\ep)-2\ep(1+2\ep))\,m\,
\Gamma[m+n,m+n-1-\ep,2\ep-m,-2\ep-n]\delta_{m>0}\,.
\end{split}
\end{equation}
\begin{equation}
\begin{split}
C^{(8)}_{k,l}=&\sum_{m=0}^k\sum_{n=0}^l
\frac{A_{k,l,m,n}}{k+l}\left[\frac{}{}(n(1-\ep)-3m\ep)\right.
\Gamma[m+n,m+n-1-\ep,1-2\ep-m,2\ep-n]\delta_{m+n>1}\\
{}&+\left.\frac{}{}\Gamma[1-2\ep,-\ep,2\ep](m/2-n\,\ep/(1-2\ep))
\delta_{m+n=1}\right]\,.
\end{split}
\end{equation}
Again, the convergence region is determined by the condition
\eqref{convergence_radius}. Similar to the previous case, at $D=4$
we reproduce from Eq. \eqref{j2_answer} the AE of the well-known result of
\cite{Usyukina1994}.

%%%%%%%%%%%%%%%%%%%%%%%%%%%%%%%%%%%%%%%%%%%%%%%%%%%%%%%%%%%%%%%%%%%%%%%%%%%%%%%%%%%%
%%%%%%%%%%%%%%%%%%%%%%%%%%%%%%%%%%%%%%%%%%%%%%%%%%%%%%%%%%%%%%%%%%%%%%%%%%%%%%%%%%%%
%%%%%%%%%%%%%%%%%%%%%%%%%%%%%%%%%%%%%%%%%%%%%%%%%%%%%%%%%%%%%%%%%%%%%%%%%%%%%%%%%%%%
%%%%%%%%%%%%%%%%%%%%%%%  Section 3   %%%%%%%%%%%%%%%%%%%%%%%%%%%%%%%%%%%%%%%%%%%%%%%
%%%%%%%%%%%%%%%%%%%%%%%%%%%%%%%%%%%%%%%%%%%%%%%%%%%%%%%%%%%%%%%%%%%%%%%%%%%%%%%%%%%%
%%%%%%%%%%%%%%%%%%%%%%%%%%%%%%%%%%%%%%%%%%%%%%%%%%%%%%%%%%%%%%%%%%%%%%%%%%%%%%%%%%%%

\section{Topology with several master integrals}

%%
%% FORWARD SCATTERING

For the case of several MIs of the same topology, when Eq.\eqref{1} turns into the system of ODE,
the formal solution of the corresponding homogeneous system is $T\exp[\int f(s)ds]$. The
asymptotical expansion of this solution can be usually found only term-by-term up to some finite
order. Our approach can be also applied in this case to find any finite-order term but, usually,
not the common term of the AE.

In this Section, we generalize our approach to the case of topology with
several master integrals. It worth noting that IBP method always leads to the
differential equations for MIs with the coefficients being rational functions
of $s$, and the point $s=\infty$ being the regular singular point. Thus,
multiplying the equations by some polynomial, we can reduce them to the
following vector form
\begin{equation}\label{de_general}
s\,P_1(1/s)\frac{\partial \mathbf{J}}{\partial
s}+\mathbb{P}_2(1/s)\mathbf{J}=\mathbf{H}\,,
\end{equation}
where $P_1(x)$ is some polynomial, $P_1(0)=1$, and $\mathbb{P}_{2}(x)$ is the polynomial matrix.
The algorithm of finding of the AE of $\mathbf{J}$ term-by-term is known from the theory of ODE.
Here we will present this algorithm in the case when the asymptotical expansion of the
inhomogeneous term is power-like,
\begin{equation}
\mathbf{H}=\sum_{\alpha\in B}\mathbf{H}(\alpha)s^{-\alpha}\,,
\end{equation}
and the following conditions hold
\begin{equation}\label{conditions}
\begin{split}
&1.\;\;\alpha_i-\alpha_j\quad \text{not integer for}\ i\neq j\,,\\
&2.\;\;\alpha_i\notin B\,,
\end{split}
\end{equation}
where $\alpha_i$ are the eigenvalues of $\mathbb{P}_{2}(0)$. In this case, we
can search for the solution of the system
\eqref{de_general} in the form of power series
\begin{equation}\label{Jseries}
\mathbf{J}=\sum_{\alpha\in A}\mathbf{C}(\alpha) s^{-\alpha}\,,
\end{equation}
where $\alpha$ is, in general, not integer, and runs over some discreet set
$A$, restricted from below. This set, as well as the coefficients
$\mathbf{C}(\alpha)$, are to be determined. It is obvious that $B\subset A$,
for convenience we can replace $B\to A$ and put $\mathbf{H}(\alpha)=0$ if
$\alpha\notin B$. Comparing the coefficient in front of equal powers of $s$, we
obtain the system of recurrence relations containing
$\mathbf{C}(\alpha)\,,\;\mathbf{C}(\alpha-1)\,,\ldots\;\mathbf{C}(\alpha-k)$
with some finite, due to the polynomial form of $P_1$ and $\mathbb{P}_{2}$,
integer $k$
\begin{equation}\label{recurrence}
[\mathbb{P}_2(\alpha^-)-P_1(\alpha^-)\alpha]\mathbf{C}(\alpha)=\mathbf{H}(\alpha)\,,
\end{equation}
where $\alpha^-$ is the shift operator, $\alpha^-f(\alpha)=f(\alpha-1)$. From Eq.
\eqref{recurrence}, we can express $\mathbf{C}(\alpha)$ as a linear combination of
$\mathbf{C}(\alpha-1)\,,\ldots\;\mathbf{C}(\alpha-k)$ and $\mathbf{H}(\alpha)$, unless $\alpha$
coincides with one of the eigenvalues of $\mathbb{P}_2(0)$. Therefore, the condition of series
breaking from below (at sufficiently small $\alpha$) results to $A=B\cup
\left\{\alpha_i,\alpha_i+1,\ldots\right\}$, where $\alpha_i$ are the eigenvalues of
$\mathbb{P}_2(0)$. If $\alpha=\alpha_i$, the coefficients $C_k(\alpha)$ can not be found from Eq.
\eqref{recurrence}. Due to the conditions \eqref{conditions}, this degenerate system remains
consistent, and we fix the coefficients $C_k(\alpha)$ by using the system and applying the boundary
conditions. It is seen, that the boundary conditions are fixed by the terms of AE of $\mathbf{J}$
proportional to $s^{-\alpha_i}$.

In the more general case, when there are $\ln s$ in the AE of $\mathbf{H}$
and/or the conditions \eqref{conditions} do not hold, we have to search the
solution in the form of the series
\eqref{Jseries} with $\mathbf{C}(\alpha)$ replaced by finite sum
$\sum_n\mathbf{C}^{(n)}(\alpha) \ln^ns$.
\begin{figure}
\centering
\[
%\fmfset{curly_len}{1mm}%
%\fmfset{thin}{0.5pt}%
%\fmfset{dash_len}{1mm}%
%\fmfset{arrow_len}{2mm}%
J_1=\hspace{-0.7cm}\raisebox{-47\unitlength}[47\unitlength]{\includegraphics{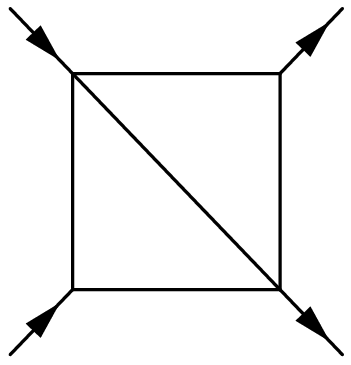}
\begin{picture}(0,0)(125,0)
\put(16.67232,13.5302){\makebox(0,0)[rb]{$p$}}%
\put(16.67232,86.4698){\makebox(0,0)[rt]{$q$}}%
\put(94.671,5.21992){\makebox(0,0)[rt]{$q$}}%
\put(103.3277,86.46982){\makebox(0,0)[lt]{$p$}}%
\end{picture}
}\qquad
%%%%
J_2=-p^2\times\hspace{-1cm}\raisebox{-47\unitlength}[47\unitlength]{\includegraphics{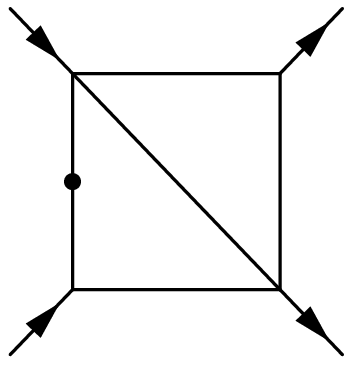}
\begin{picture}(0,0)(125,0)
\put(16.67232,13.5302){\makebox(0,0)[rb]{$p$}}%
\put(16.67232,86.4698){\makebox(0,0)[rt]{$q$}}%
\put(94.671,5.21992){\makebox(0,0)[rt]{$q$}}%
\put(103.3277,86.46982){\makebox(0,0)[lt]{$p$}}%
\end{picture}}\qquad
%%%%
J_3=p\cdot q
 \times\hspace{-1cm}\raisebox{-47\unitlength}[47\unitlength]{\includegraphics{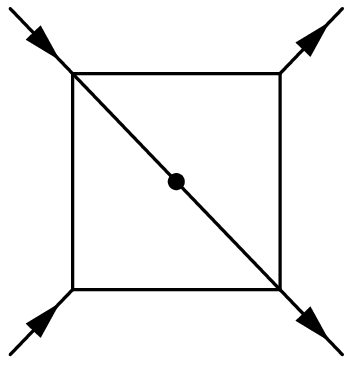}
 \begin{picture}(0,0)(125,0)
\put(16.67232,13.5302){\makebox(0,0)[rb]{$p$}}%
\put(16.67232,86.4698){\makebox(0,0)[rt]{$q$}}%
\put(94.671,5.21992){\makebox(0,0)[rt]{$q$}}%
\put(103.3277,86.46982){\makebox(0,0)[lt]{$p$}}%
\end{picture}}
\]
\hfill a \hfill\hfill b\hfill\hfill c\hfill \caption{"Crossed--box" topology.} \label{Diamond}
\end{figure}

As an example, we consider ``crossed--box'' topology in forward kinematics (see
Fig.\ref{Diamond}) with  three master integrals. In particular, these MIs
appear in calculations of radiative corrections to cross sections and of impact
factors in QCD. We choose the MIs as shown in Fig.\ref{Diamond}, i.e.,
\begin{align}
J_1&=\frac{1}{(2\pi)^{2D}}\int\frac{d^D r d^D l}{r^2
    l^2(r+p)^2(l+p)^2(l-r+q)^2}\,,\label{J1}\\
J_2&=-p^2\frac{1}{(2\pi)^{2D}}\int\frac{d^D r d^D l}{[r^2]^2
l^2(r+p)^2(l+p)^2(l-r+q)^2}\,,\label{J2}\\
J_3&=(p\cdot q)\frac{1}{(2\pi)^{2D}}\int\frac{d^D r d^D l}{r^2
    l^2(r+p)^2(l+p)^2[(l-r+q)^2]^2}\,.\label{J3}
\end{align}

The master integrals \eqref{J1}--\eqref{J3} obey the following system of
differential equations
\begin{align}\label{de31}
\lambda^2\frac{\partial J_1}{\partial s}&=-2y \epsilon J_1-y
J_2+\left(q^2+y\right)
   J_3+H_1\,,\\
\begin{split}\lambda ^2u y\frac{\partial J_2}{\partial s}&= \epsilon
(1-6\epsilon)y p^2\frac{3u+q^2-p^2}{4} J_1 -\frac{y}{2}\left(u y+2
\left(p^4+4 p^2(q^2 - y)-q^2 y\right)
\epsilon\right) J_2\\
&-\epsilon\left(\lambda ^2s-y\left(p^2-q^2\right)
   \left(u-q^2\right)\right) J_3+ H_2
\,,\end{split}\label{de32}\\
\begin{split}
2\lambda ^2 s u y\frac{\partial J_3}{\partial s}&= y p^2 \epsilon (1-6 \epsilon
)\left(\left(p^2-2 q^2\right)
   \left(p^2-q^2\right)-6\lambda ^2\right) J_1\\
   &{}+2\epsilon y\left(
    2\lambda ^2(4 p^2+q^2)-p^2 \left(p^2-2 q^2\right)
   \left(p^2-q^2\right)\right) J_2\\
   &+\left(-\left(p^2-q^2\right) \epsilon  p^2 y s-4\lambda ^2u y \epsilon
   -2\lambda ^2s u\epsilon+suy(3 \left(p^2+2 y\right) \epsilon -y)\right) J_3
+ H_3\,,
\end{split}\label{de33}
\end{align}
where $u=(p-q)^2=2p^2+2q^2-s$ and $H_i$ are expressed in terms of MIs of
subtopologies of ``crossed--box'' topology and presented in the Appendix. In
this case
\begin{equation}
\mathbb{P}_{2}(0)=\left(
                     \begin{array}{lll}
                       4\epsilon & 2 & -2 \\
                       0         & 1 & -2\epsilon \\
                       0         & 0 & 1-2\epsilon \\
                     \end{array}
                   \right)\,,
\end{equation}
the eigenvalues being $\alpha_{1,2,3}=4\epsilon,\,1,\,1-2\epsilon$. Therefore,
we have to fix the three terms of AE of $J_1,\,J_2,\,J_3$ proportional to
$s^{-4\epsilon}\,,s^{-1}\,,s^{-1+2\epsilon}$, for example,
$C_1(4\epsilon),\,C_1(1),\,C_2(1-2\epsilon)$.

First, we calculate the asymptotics of $J_1$. The Feynman parametrization of
$J_1$ reads
\begin{equation}\label{fp3}
J_1=
\frac{\Gamma[1+2\epsilon]}{(4\pi)^D}\int_0^\infty\frac{\prod_{i=1}^5dx_i\,
\delta(1-x_{12345})(x_{12}x_{34}+x_5x_{1234})^{-1+3\epsilon}}{(-p^2
(x_5x_{14}x_{23}+x_2x_3x_{14}+x_1x_4x_{23})-2y
x_5(x_2x_4-x_1x_3)-q^2x_5x_{12}x_{34})^{1+2\epsilon}}\,.
\end{equation}
Again, we assume $\epsilon>0$. It is easy to see that there is no contribution
$\propto s^{-4\epsilon}$, therefore, $C_1(4\epsilon)=0$. The main asymptotics
$\propto s^{-1}$ is given by the region
\begin{equation}\label{region3}
0\leqslant x_5\ll x_1 \sim x_2\sim x_3\sim x_4 \leqslant 1\,.
\end{equation}
The integral should be treated with care. Assuming that $-p^2,-q^2>0$ and $\mathrm{Im} (s)\neq0$,
we obtain
\begin{equation}
J_1\stackrel {s\to\infty}\longrightarrow \frac{C_1(1)}{s}=\frac{i\sigma
\pi}{(4\pi)^D}\frac{(-p^2)^{-2\epsilon}}{s}
\frac{\Gamma[\epsilon]^2\Gamma[1-2\epsilon]^2}{\Gamma[2-4\epsilon]}\,,
\end{equation}
where $\sigma=\mathrm{sign}( \mathrm{Im}\  s)=(-i/\pi)\lim_{s\to \infty} \ln (s/u)$.

Now we calculate the asymptotics of $J_2$. When $\epsilon>0$ the integral in
Eq. \eqref{J2} is infrared divergent. This divergence can be eliminated by
subtracting from $J_2$ the scaleless integral
\begin{equation}
-p^2\frac{1}{(2\pi)^{2D}}\int\frac{d^D r d^D l}{(r^2)^2
l^2p^2(l+p)^2(l+q)^2}\,.
\end{equation}
This subtraction determines the analytical continuation of $J_2$ to the region $D<4$ ($\ep >0$).
After the Feynman parametrization, we have
\begin{equation}
\begin{split}
J_2&=p^2\frac{\Gamma[2+2\epsilon]}{(4\pi)^D}\int\limits_0^\infty\prod_{i=1}^5dx_i\,
\delta(1-x_{12345})\\
&\times\left[\frac{x_1(x_{12}x_{34}+x_5x_{1234})^{3\epsilon}}{(-p^2
(x_5x_{14}x_{23}+x_2x_3x_{14}+x_1x_4x_{23})-2y
x_5(x_2x_4-x_1x_3)-q^2x_5x_{12}x_{34})^{2+2\epsilon}}\right.\\
&{}-\left.\frac{x_1^{-1+\epsilon}(x_{345})^{3\epsilon}}{(-p^2
(x_{23}x_{45}+x_2x_3)+2y x_3x_5-q^2x_5x_{34})^{2+2\epsilon}} \right]
\,.
\end{split}
\end{equation}
The main asymptotics is given by the region
\begin{equation}\label{region4}
0\leqslant x_2\sim x_3 \sim x_5\ll x_1\sim x_4 \leqslant 1\,.
\end{equation}
The final result reads
\begin{equation}
J_2\stackrel {s\to\infty}\longrightarrow
\frac{C_2(1-2\epsilon)}{s^{1-2\epsilon}}=\frac{\Gamma[2\epsilon]}{(4\pi)^D}
\frac{\pi^2(-p^2)^{-2\epsilon}(-q^2)^{-2\epsilon}}{\sin(2\pi\epsilon)\sin(\pi\epsilon)}
\left[s^{-1+2\epsilon}-(-s)^{-1+2\epsilon}\right]\,.
\end{equation}

The AE of inhomogeneous terms determines the set $B$,
\begin{equation}
B=\bigcup_{k=1}^{\infty}\{k-\epsilon,k,k+\epsilon,k+2\epsilon\}\,.
\end{equation}
As we can see, the second condition in Eq. \eqref{conditions} is not satisfied.
However, we have checked that the system of recurrence relations remains
consistent and, therefore, we can search for the solution in the form
\eqref{Jseries} with the set $A$ defined as follows
\begin{equation}
A=\bigcup_{k=1}^{\infty}\{k-2\epsilon,k-\epsilon,k,k+\epsilon,k+2\epsilon\}.
\end{equation}
The set $\{4\epsilon,1+4\epsilon,\ldots\}$ is not included because
$C_1(4\epsilon)=0$. To save space, we do not present the explicit form of
recurrence relations which are readily obtained from
Eqs.\eqref{de31}--\eqref{de33}. One of the consequences of these relations is
that ${C_3(1-2\epsilon)}={C_2(1-2\epsilon)}$ and, therefore, the main
asymptotics of $J_3$  reads
\begin{equation}
J_3\stackrel {s\to\infty}\longrightarrow \frac{\Gamma[2\epsilon]}{(4\pi)^D}
\frac{\pi^2(-p^2)^{-2\epsilon}(-q^2)^{-2\epsilon}}{\sin(2\pi\epsilon)\sin(\pi\epsilon)}
\left[s^{-1+2\epsilon}-(-s)^{-1+2\epsilon}\right]\,.
\end{equation}

\section{Conclusion}

In this paper we present the approach to the calculation of the dimensionally regularized multiloop
integrals. This approach is based on the differential equations method and asymptotical expansion.
As a demonstration of our method, we have calculated the asymptotical expansions \eqref{j1_answer}
and \eqref{j2_answer} of the master integral, depicted in Fig.\ref{Marquee} in two different
kinematical regions. Moreover, we have stressed that these asymptotical expansions have a finite
convergence radius determined by Eq.\eqref{convergence_radius}, thus being the power series
accessible for the numerical calculations. It worth noting that this master integral was known only
in $O(\epsilon)$ order so far \cite{Birthwright2004} and this expansion was much more lengthier
than our exact representation. Our results can be easily expanded up to any order of $\epsilon$. We
have checked that in the limit $\epsilon\to 0$ our result coincides with the well-known result of
Ref. \cite{Usyukina1994}. The result for special kinematics when one or two external momenta are on
mass shell ($p^2=0$ and/or $q^2=0$) can be readily obtain from Eqs.
\eqref{j1_answer}-\eqref{j2_answer}. The case when a topology contains several master integrals
has also been considered. We present an algorithm of the term-by-term calculation of the
asymptotical expansion in this case. The "crossed--box" topology with three master integrals shown
in Fig.\ref{Diamond} is analyzed in detail.

\section{Appendix}
The master integrals appearing in the inhomogeneous parts of differential equations \eqref{de1},
\eqref{de2}, \eqref{de31}--\eqref{de33} are conveniently expressed via the one--loop off--shell
triangle with arbitrary powers of denominators
\begin{equation}
    J(\mu,\nu,\rho)=\frac{1}{(2\pi)^D}\int\frac{d^D r}{[-r^2]^\mu
    [-(p-r)^2]^\nu[-(q+r)^2]^\rho}\,.
\end{equation}
The well-known result \cite{Boos1991} for this integral reads
\begin{equation}
\begin{split}
& J(\mu,\nu,\rho)=\frac{i}{(4\pi)^{D/2}}
\frac{\pi^2\csc(D/2-\mu-\rho)\csc(D/2-\mu-\nu)}{\Gamma[\mu,\nu,\rho,D-\mu-\nu-\rho]}
 \sum_{n,m=0}^\infty\frac{(p^2/s)^n(q^2/s)^m}{n!\;m!}\times\mbox{}\\
 &\left((-s)^{\mu-D/2}(-p^2)^{D/2-\mu-\nu }
 (-q^2)^{D/2-\mu -\rho }
\Gamma\left[\substack{\displaystyle D-\mu
-\nu-\rho+n+m,D/2-\mu+n+m\\ \displaystyle
 D/2-\mu -\nu+1+n,D/2-\mu -\rho+1+m}\right]\right.\\
&\mbox{}-(-s)^{-\rho}(-p^2)^{D/2-\mu -\nu }
\Gamma\left[\substack{\displaystyle\rho+n+m,D/2-\nu+n+m\\
\displaystyle D/2-\mu-\nu+1+n,\mu+\rho-D/2+1+m}\right]\\
&\mbox{}-(-s)^{-\nu}(-q^2)^{D/2-\mu -\rho }
\Gamma\left[\substack{\displaystyle\nu+n+m,D/2-\rho+n+m\\
\displaystyle
\mu+\nu-D/2+1+n,D/2-\mu-\rho+1+m}\right]\\
&\mbox{}+\left.(-s)^{D/2-\mu -\nu -\rho }
\Gamma\left[\substack{\displaystyle \mu+n+m ,\mu+\nu+\rho-D/2+n+m\\
\displaystyle\mu+\nu-D/2+n +1,\mu+\rho-D/2+1+m}\right] \right)\,.
\end{split}
\end{equation}
Introducing the notation
\begin{equation}\label{bubble1}
J(\mu,\nu)=\frac{i}{(4\pi)^{D/2}}
\frac{\Gamma[\mu+\nu-D/2,D/2-\mu,D/2-\nu]}{\Gamma[\mu,\nu,D-\mu-\nu]}\,,
\end{equation}
we obtain for the master integrals appearing in Eqs. \eqref{de1}, \eqref{de2}
\begin{equation}
\begin{split}
F_n(s,p^2,q^2)&=(-1)^{n+1}J(1,n) J(1+n-D/2,1,1)\,,\\
F_n(p^2,q^2,s)&=(-1)^{n+1}J(1,n) J(1,1,1+n-D/2)\,,\\
F_n(q^2,p^2,s)&=(-1)^{n+1}J(1,n) J(1,1+n-D/2,1)\,,\\
G(s)&=J(1,1) (-s)^{D/2-2}\,.
\end{split}
\end{equation}
The inhomogeneous part of the system \eqref{de31}--\eqref{de32} is expressed as
follows
\begin{equation}
H_1=\left(q^2 + y\right) T_7 - \left(q^2 + 2 y\right) T_8\,,
\end{equation}
\begin{equation}
\begin{split}
H_2&=(q^2-p^2- u) y (1 -
              2 \epsilon )^2 T_1/4 + (q^2-p^2+u) y (1 - 3 \epsilon ) (2 - 3 \epsilon ) (1 -
            2 \epsilon ) T_2/(4 q^2 \epsilon)\\
            {}&- y (1 -
            3 \epsilon ) (2 - 3 \epsilon ) (1 -
            2 \epsilon ) T_3/(2 \epsilon ) + (q^2-p^2  +
        u) y (1 - 3 \epsilon ) (1 -
            2 \epsilon ) T_5/4\\
            {}& + (p^2 -
        q^2) y (1 - 3 \epsilon ) (1 - 2 \epsilon ) T_6/2 + ( q^4-p^4  + u (4 p^2 - q^2 +
              3 y)/2) \epsilon  y T_7\\
              {}& - (3 (q^2 -p^2
                 p^2 + (7 p^2 + q^2) u +
          4 u y)\epsilon  y T_8/2\,,
\end{split}
\end{equation}
\begin{equation}
\begin{split}
H_3&=(q^4-p^4  -
        s u) y (1 -
              2 \epsilon )^2 T_1 /2- (q^4-p^4  +
        s u) y (1 - 3 \epsilon ) (2 - 3 \epsilon ) (1 -
            2 \epsilon ) T_2/(q^2 \epsilon )\\
            {}& + s y (1 -
            3 \epsilon ) (2 - 3 \epsilon ) (1 -
            2 \epsilon ) (s T_3+uT_4)/\epsilon + (p^4 - q^4 +
        s u) y (1 - 3 \epsilon ) (1 -
            2 \epsilon )(T_5+T_6)\\
                     {}& +
  s y ( q^4-p^4  + (3 p^2 - q^2) u +
          2 u y) \epsilon  T_7 \\
          {}&-
  y (s u (9 p^2 + 4 q^2 +
                2 y) -(p^4 - q^4) (5 p^2 - 4 q^2 +
                2 y)) \epsilon  T_8\,,
\end{split}
\end{equation}
where
\begin{align}
&T_1=G^2(p^2)\,,&&T_5=F_1(p^2,q^2,u)\,,\nonumber\\
&T_2=J(1,1)J(1,2-D/2)(-q^2)^{D-3}\,,&&T_6=F_1(p^2,q^2,s)\,,\nonumber\\
&T_3=J(1,1)J(1,2-D/2)(-u^2)^{D-3}\,,&&T_7=F_2(p^2,q^2,s)\,,\\
&T_4=J(1,1)J(1,2-D/2)(-s^2)^{D-3}\,,&&T_8=F_2(p^2,q^2,u)\,.\nonumber
\end{align}

%\bibliography{Refs}

\end{document}